# CONTRIBUTIONS TO THE DEVELOPMENT AND IMPROVEMENT OF A REGULATORY AND PRE-REGULATORY DIGITALLY SYSTEM FOR THE TOOLS WITHIN FLEXIBLE FABRICATION SYSTEMS

**Viorel PUTZ[1] and Mihai V. PUTZ[2]**

[1] Polytechnic University of Timisoara, Department of Manufacturing Engineering, Avenue M. Viteazul No. 1, RO-300222, Timisoara, Romania, e-mail: vputz@eng.utt.ro
[2] West University of Timisoara, Department of Chemistry, Avenue Pestalozzi, No.16, RO-300115, Timisoara, Romania, e-mail: mv.putz@home.ro

## Abstract

The paper reports the obtained results for the projection and realization of a digitally system aiming to assist the equipment for a regulatory and pre-regulatory tools and holding tools within the flexible fabrication systems (FFS). Moreover, based on the present results, the same methodology can be applied for assisting tools from the point of view of their integrity and to wear compensation in the FFS framework.

# 1. Conceptual Considerations about the State of Art of Data Acquisition Systems (DAS)

The previous researches on the *flexible fabrication systems* (FFS) have revealed the necessity to adopt a *digitally managing system* production depending on the particular considered case. [1-5]

The actual managing systems are based on the hierarchical systems, equipped with a computing net as well providing a distributed intelligence. [3,5] This way, the developing of the standard interfaces coupled with the implementation of the modularly construction principle, and together with data acquisition modeling, should competitively improve the technologic FFS. In this context, the principal functions of the managing FFS rely on the following list of indicators:

- *automatically dispatcher* of the working process;

- *functional harmonization* of the working system with the processes and activities engaged on the assembly section and the supplying services, respectively;

- *assuring of the functional autonomy of the tool-machine* and of the auxiliary sub-systems through their fitting with microprocessors and micro-computers;

- *implementation and developing* of the fabrication systems of converting type from analogical to digitally and mechanical response, CAD-CAM, assuring therefore an unique flux of information from the projection to the final finite piece.

For any managing system, there is assumed that its meaning is to elaborate the right commands for the whished working process. However, controlling the technical systems implies the use of complex computing systems that, nevertheless, introduce inherent modeling indeterminacy and uncertainty. [2] These last factors become critical in managing the real systems pointing out the difficulties of dealing with complex systems. Therefore, approaching the complex systems states as an essential challenge in the frame of the measure and control system theories. Up to now, there are some clarified aspects regarding the automat controlling of the complex systems, for instance:

- *keeping the human supervising* as a central link among input-output of the complex system;
- *semiautomatic control* of the flowing processes;
- *automat managing* of the integral system.

At least in Romania, the actual trend is to maintain the human supervising as the main decision factor. Instead, in a hi-technological economies the trend is more orientated to the flexible production, mainly based on the digitally assisted process. The last attitude is founded on minimum two considerations, namely:

- binary computers are excellent supports for decisions;
- not all stages of a process requires the definitely human involvements.

The data acquisition systems (DAS) are localized on the reaction circuit of a command-control system. [2, 3] Basically, DAS carry out for the next functions:

- to convert the input signals from analog-to-numeric data;
- to process the conversion of resulted numerical data;
- to store the numerical data in files;
- to mould and transmit the output data.

The natural consequence of the above state of art in a technological production process is to introduce the concept of the *intelligent* or *digitally system*. One strong argument of this "intelligent" attribute for digitally evolved systems calls their *functional autonomy*. The development and implementing of such advanced systems in the fabrication programs constitutes a major research direction for the technological progress.

# 2. Analysis of the Constructive Solution of the Data Acquisition System (DAS) from a Flexible Fabrication Process

The usefulness of computers in applications is grounded on automatically procession of data that further inform on the nature of the concerned application.

However, in many applications, the data to be numerical analyzed are time-dependent, being correlated with the evolving physically indices associated with the process. Therefore, for a proper numerical analyze, the physical data are firstly transformed (or converted) into analogical signals and then in numerical signals, throughout translators and acquisition systems, respectively.

The produced operations of the numerical working systems, for instance the system MICROEL 01 sketched in Figure 1, on the acquisition numerical signals can be grouped as: filtering, frequency range representation, classification, identification, etc. [6]

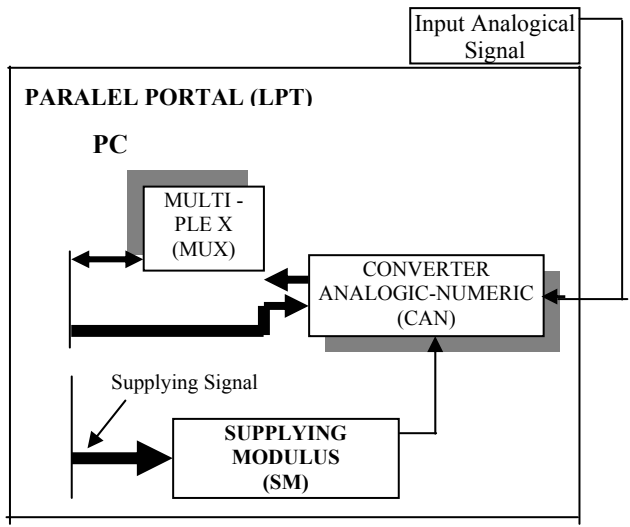

**Figure 1.** *The block scheme of the MICROEL 01-DAS interface (see the text for details).*

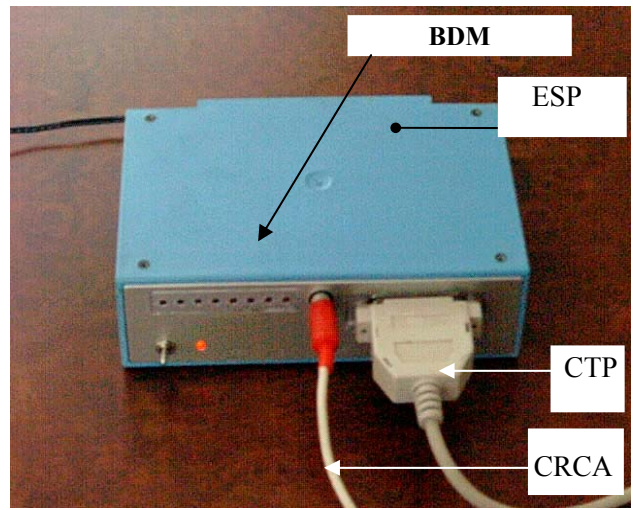

**Figure 2.** *Outside view of the MICROEL 01-DAS (see the text for details).*

Also the numerical signals resulted from processing can be back converted into analogical ones. Therefore, the working information can be contained both as numerical or analogical signals, being further suited for memory, reading, communicating or control.

The physical processes are characterized by quantities which can be measured performing their transformation into electrical (analogical) signals by means of translators. Afterwards, the signals can be worked out both by analogical or numerical techniques.

Firstly, numerical analysis requires the additional transformation of the analogical signals in numerical ones by using a data acquisition system-DAS. Then, numerical data are recorded by noting, at definite intervals of time, the "shape" values (amplitudes for instance) of the converted analogical signals. Thus, the basic components of DAS are the pattern and memorizing circuits together with the converters of analogical-numerical (CAN) signals.

An original DAS scheme is presented in Figure 1 as the **"MICROEL 01"** data acquisition interface. [6] It is completely modular, each modulus being separately realized and optimized according with some guidelines:
- *safely* function;
- minimum error factor (under 1%);
- *real time analogical-numerical conversion*;
- *binary display* of conversion;
- *high grade protection*, with a ionization potential IP=50V of the electrical insulation, between the supplying tension ( 220V cca.) and the working modular voltage;
- *the stability* conversion at the environment temperature variations;
- *trendy design*, see Figure 2;
- *standard connectivity* with a computer station (PC, LPT).

The MICROEL 01 is aiming to an analogical-digitally conversion for an input voltage in the range 0 – 5V cca., with an output of 8 bites. Practically, this conversions is placed in the principal modulus (CAN) of the apparatus, see Figure 1. The multiplex block (MUX) has the role to combine the lines furnished by CAN. The reading data from CAN are performed on the 4 state lines (SEL, P.END, ACK, BUSY) of the parallel portal.

For the all CAN-portal 4 lines the recording of the values encoded in the 8 bites is done in two times: firstly is decoded the inferior octet, then the superior one. The multiplex block is implemented with a 74HC244 circuit allowing 8 and 4 signals for the input and output, respectively connections (8:4).

The CAN device realizes its conversion function within a certain time period, leading thus with a discretized signal, represented by values of the analogical input signal collected at well defined time intervals. The "organic" conversion of an analogical signal into a numerical one is principally produced as follows:

- *the discretization (the pattern) stage* of the input signal is realized at the definite time periods, such that the time range is quantized in the multiples of the so called "pattern steps" (or pattern frequency), accompanied by the memorization of the signal values (amplitudes) at those temporal steps;
- *the conversion (the digitally) stage* regards the transcription of the recorded pattern value into a binary number, or, as is the present case, into a 8 binary digits (or bites) number.

The binary displayed modulus (BDM) consists therefore from 8 LED units connected with the conversion modulus CAN such that to provide a real time posting of the binary converted information.

The device just presented means the constructive solution for the DAS of a process, leading open the further possibilities of its connection through direct intelligent lines also with other periphery digitally equipments. This way, an adaptive command of DAS can be applied in managing of the regulatory and pre-regulatory tools and holding tools within flexible fabrication processes. Such an example, in terms of wear knife toll compensation, through its radial moving relatively to the working piece, is exposed in the next section.

## 3. Basic Solution for a Compensation Subsystem of the Knife Tool Wear

On behalf of the FFS, both the wear as well the regulatory error can be automatically compensated. [1, 2, 4]. Regarding this aspect, in the specialized literature, there are largely presented and discussed various solutions and the active control systems for a series of technological cutting processes. In this frame, we can develop another wear compensation solution for the working tool by using the previous proposed DAS solution.

In Figure 3, is presented the control solution with direct contact on the edge tool, through a translator which continuously emits a signal and that, in the normal conditions, has to fit with the edging working tool SA.

When the toll is absent, because the wear advancing process, the emitted signal of the translator element is amplified in A, then will be applied on the impulses generator block (FI) that supplies, with an adjustable frequency, the executable element from the electromagnet E. As a consequence, the core movement of the electromagnet E will linearly train the gang arm BC, which, in its turn, will move the teeth of the gang wheel RC, producing a certain revolution that imposes also the rotation of the snail's device MRMM. Together with the snail's wheel, also the conducting screw SC is moving, which, through the screw nut (not figured), will produce the finally movement of the holding-knife PC with an equal rate of the advancing wear of the working tool SA. Because the holding-tools are moving on guide with radial rolling is compulsory to stuck it in the working position at the final of the compensation process. Such a blockage is provided through a pneumatic system, with the linear pneumatic engines that will activate the brakes that keep in the working position the holding-knife PC.

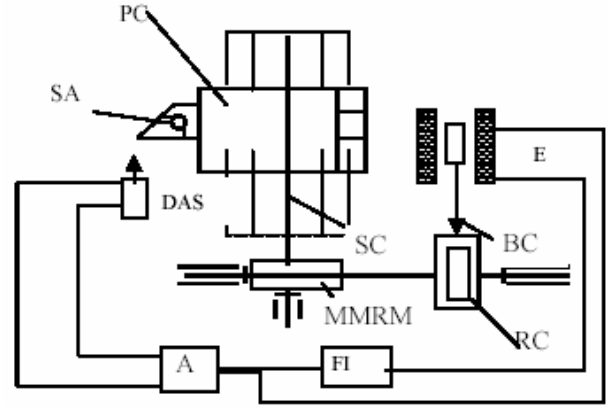

**Figure 3.** *Basic solution scheme for the wear compensation subsystem (see the text for details).*

Respecting the available solutions in literature, the present one displays a more unitary and precision character that can be further improved once a step by step engine is coupled with the holding knife device.

## 4. Conclusions

In the framework of flexible fabrication processes there is reported and analyzed an original constructive solution both for data acquisition systems as well for compensation subsystem of the knife tool wear.

The results are valuable, being tested on the level of a working tool center, allowing future integration of the present achievements as flexible sub-systems cells of a more complex flexible fabrication system.

As an important part of flexible production process, the compensation operation of wear's tools has to be rigorously and economically integrated in the strategy and costs of any piece fabrications.

The necessary research on the way of implementing the data acquisition systems for a net of compensation flexible processes is therefore possible, and is actually in progress.